\documentclass[twocolumn,english]{IEEEtran}
\usepackage[T1]{fontenc}
\usepackage{color}
\usepackage{babel}
\usepackage{textcomp}
\usepackage{amsmath}
\usepackage{amssymb}
\usepackage{graphicx}
\usepackage[unicode=true,
 bookmarks=true,bookmarksnumbered=true,bookmarksopen=true,bookmarksopenlevel=1,
 breaklinks=false,pdfborder={0 0 0},backref=false,colorlinks=true]
 {hyperref}
\hypersetup{pdftitle={Your Title},
 pdfauthor={Your Name},
 pdfpagelayout=OneColumn, pdfnewwindow=true, pdfstartview=XYZ, plainpages=falsem,citecolor=blue}

\makeatletter

\newcommand{\lyxdot}{.}

 \let\oldforeign@language\foreign@language
 \DeclareRobustCommand{\foreign@language}[1]{%
   \lowercase{\oldforeign@language{#1}}}

\ifCLASSOPTIONcompsoc
\usepackage[caption=false,font=normalsize,labelfont=sf,textfont=sf]{subfig}
\else
\usepackage[caption=false,font=footnotesize]{subfig}
\fi

\usepackage{lineno}

\makeatother

\begin{document}

\title{Control and generation of localized pulses in passively mode-locked
semiconductor lasers }

\author{M.~Marconi, J. Javaloyes, P. Camelin, D. Chaparro, S. Balle, and~M.
Giudici~\IEEEmembership{Member,~IEEE}.%
\thanks{M. Marconi, P. Camelin and M. Giudici are with the Institut Non Linéaire
de Nice, Université de Nice Sophia Antipolis - Centre National de
la Recherche Scientifique, 1361 route des lucioles, F-06560 Valbonne,
France, e-mails: \protect\href{http://mathias.marconi@inln.cnrs.fr}{mathias.marconi@inln.cnrs.fr},
\protect\href{http://patrice.camelin@inln.cnrs.fr}{patrice.camelin@inln.cnrs.fr}
and \protect\href{http://massimo.giudici@inln.cnrs.fr}{massimo.giudici@inln.cnrs.fr}.%
}%
\thanks{D. Chaparro and J. Javaloyes are with the Departament de Fisica, Universitat
de les Illes Baleares, C/ Valldemossa, km 7.5, E-07122 Palma de Mallorca,
Spain, e-mail: \protect\href{http://daniel.chaparro@uib.es}{daniel.chaparro@uib.es}
and \protect\href{http://julien.javaloyes@uib.es}{julien.javaloyes@uib.es}%
}%
\thanks{S. Balle is with the Institut Mediterrani d'Estudis Avançats, CSIC-UIB,
E-07071 Palma de Mallorca, Spain, e-mail: \protect\href{http://salvador@imedea.uib-csic.es}{salvador@imedea.uib-csic.es}%
}}

\markboth{Journal of Selected Topics in Quantum Electronics}{Marconi \MakeLowercase{\emph{et al.}}:
Control and Generation of Independent pulses}
\maketitle
\begin{abstract}
We show experimentally and theoretically that localized pulses can
be generated from an electrically biased $200\,\mu$m multi-transverse
mode Vertical-Cavity Surface-Emitting Laser. The device is passively
mode-locked using optical feedback from a distant Resonant Saturable
Absorber Mirror and it is operated below threshold. We observe multistability
between the off solution and a large variety of pulsating solutions
with different number and arrangements of pulses per round-trip, thus
indicating that the mode-locked pulses are localized, i.e. mutually
independent. We show that a modulation of the bias current allows
controlling the number of the pulses travelling within the cavity,
thus suggesting that our system can be operated as an arbitrary pattern
generator of 10~ps pulses and 1~W peak power.\end{abstract}
\begin{IEEEkeywords}
Mode-Locking, Broad-Area Lasers, VCSELs
\end{IEEEkeywords}

\section{Introduction}

\IEEEPARstart{M}ode-locking (ML) is a fascinating phenomenon that
allows the generation of ultrashort pulses from a laser \cite{haus00rev}
and that is still a subject of intense research. Passive ML (PML)
is arguably the most successful approach and it is achieved by combining
two elements, a laser amplifier which provides gain and a saturable
absorber (SA) acting as a pulse shortening element, see \cite{haus00rev}
for a review. Pulsed emission with a fundamental period corresponding
to the cavity round-trip time arises from the different dynamical
properties of the SA and the amplifier, which open a short window
for amplification around the pulse \cite{haus75f,haus75s}. ML has
led to the shortest and most intense optical pulses ever generated
and pulses in the femtosecond range are produced by dye \cite{FSY-JQE-83}
and solid state lasers \cite{keller96}. Large output powers in the
Watt range are commonly achieved from coupled VECSELs-SESAM \cite{haring02}
configurations. On the other hand, the large gain of semiconductor
materials allows building sub-millimeter Monolithic PML lasers. The
round-trip of such short devices is typically of the order of $10\,$
ps and they can therefore reach repetition frequencies of several
tens of GHz. They have also the advantage of being compact, low cost
and adaptable to many cavity geometries \cite{avrutin00}, yet their
power is in the milli-Watt range.

In spite of the research efforts dedicated to the general understanding
of the basic issues, several aspects of PML still represent a scientific
challenge. For instance, while the breadth of the active medium gain
curve is well known to govern the pulse-width, it was only very recently
understood that the SA also provides a strongly asymmetric and non
linear filtering. Such an usually overlooked effect was recently proven
to be the key mechanism explaining wavelength instabilities in PML
lasers \cite{SJM-PJ-11,TJA-JSTQE-13}. In addition, recent studies
indicated that the optimal position of the SA in the cavity was not
following intuitive rules \cite{JB-OL-11}. Similarly, it was commonly
thought until recently that PML semiconductor lasers can not operate
at low repetition rates due to the fast recovery time of their material
gain ($\tau_{g}\sim1\,$ ns) which should limit them to high repetition
frequencies ($\gtrsim1\,$ GHz). Too long cavities result in the so-called
regime of harmonic mode-locking in which several pulses circulate
in the cavity, according to the background stability criterion \cite{haus00rev}.
Until recently, the record of the lowest frequency PML was experimentally
attained around $\sim300\,$MHz \cite{MTP-JSTQE-11,CYN-JSTQE-11,BSK-EL-13}.

We have recently shown that\emph{,} in the limit of cavity round-trip
much longer than the gain recovering time, mode-locked pulses may
coexist with the zero intensity background and can be interpreted
as\emph{ temporal Localized Structures} (LS) \cite{MJB-PRL-14}. Localized
pulses can be independently addressed and used as elementary bits
of information, hence the cavity can be used as an all-optical buffer,
as shown in \cite{LCK-NAP-10}. A single localized pulse can be activated
within the cavity, independently from the cavity size, thus leading
to arbitrary low repetition rates which allowed establishing a new
low frequency record of $65.5\,$MHz \cite{MJB-PRL-14}. The possibility
of addressing individually localized pulses opens interesting possibilities
for the optical generation of arbitrary trains of narrow pulses, which
has a large number of potential applications in different domains,
e.g. time-resolved spectroscopy, pump-probe sensing of material properties,
generation of frequency combs, Optical Code Division Multiple Access
Communication Networks \cite{ocdma} and LIDAR \cite{lidar,lidar2}.

In this manuscript, we show how localized pulses form in a passively
mode-locked semiconductor broad-area VCSEL coupled to a resonant saturable
absorber mirror (RSAM). The VCSEL bias current is used as a control
parameter for addressing localized pulses and for generating patterns
of closely packed narrow pulses. The experimental evidences and theoretical
analysis pave the path towards an optical arbitrary pattern generator
of short light pulses.

\section{Experimental Results}

\subsection{Experimental setup}

\begin{figure}[h]
\centering{}\includegraphics[bb=0bp 0bp 1580bp 1772bp,clip,width=0.37\textwidth]{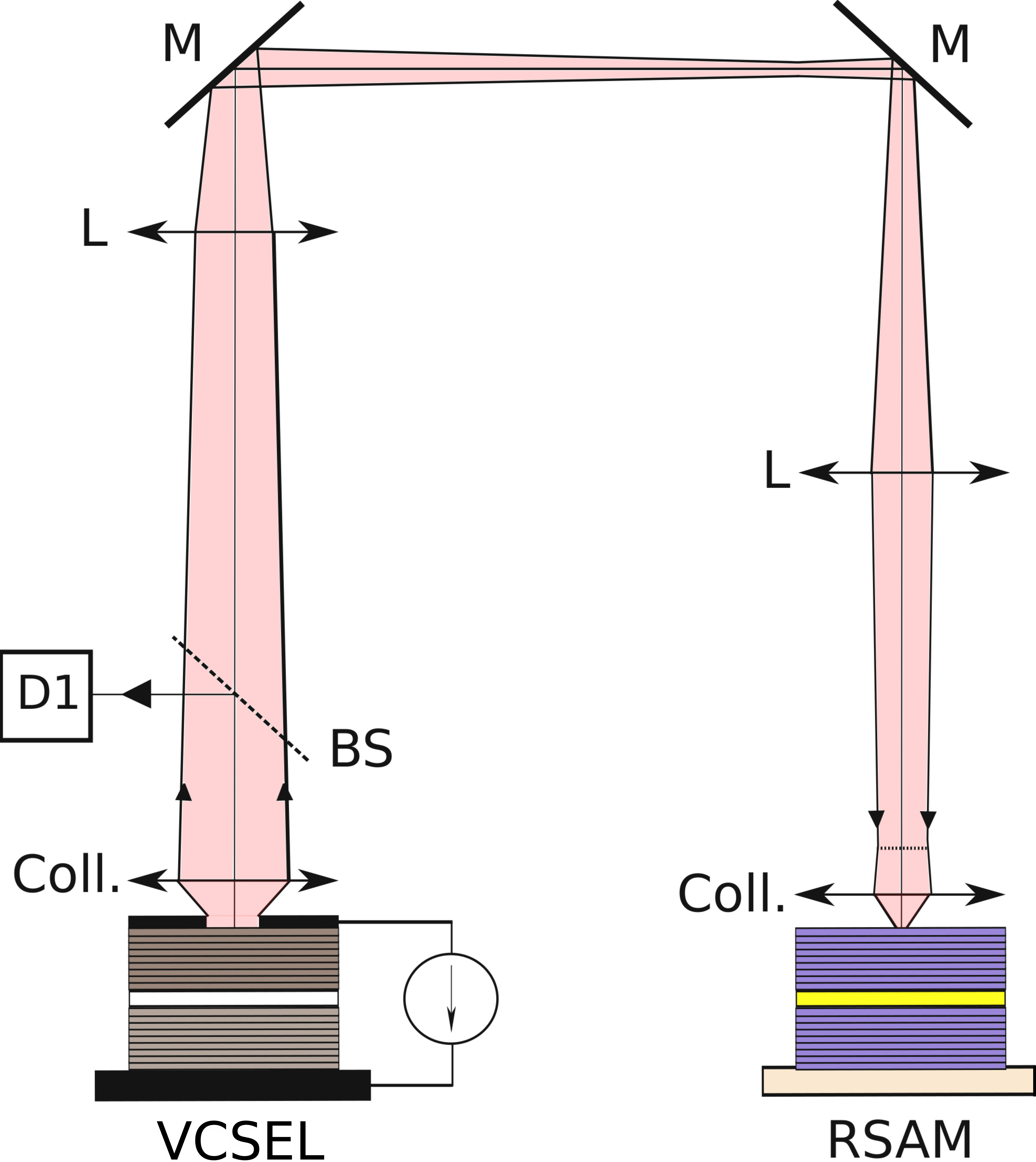}
\caption{Experimental Set-up: Temperature-stabilized VCSEL and RSAM. Coll.:
Aspheric Lens, BS : Beam Splitter, M: Mirror and D1: Detector and
CCD cameras.\label{setup}}
\end{figure}

Our VCSEL is a $980\,$nm device manufactured by ULM Photonics \cite{701502}.
Its standalone threshold current ($J_{st}$) is about $380\,$mA,
though emission is localized only at the periphery of the device up
to $J=850\,$mA, after which roll-off starts to occur. The $980\,$nm
RSAM (BaTop Gmbh) has a transverse dimension of $4\times4\,$mm$^{2}$
and it exhibits a low unsaturated reflectivity of $1\,\%$ that increases
up to $60\,\%$ when saturated. The RSAM saturation fluence is $15\,\mu$J.cm$^{-2}$.
These values are obtained at the RSAM resonant wavelength which can
be thermally tuned over $3\,$nm (between $T_{1}=10\,\text{\textdegree C}$
and $T_{2}=50\,\text{\textdegree C}$). The Full Width at Half Maximum
(FWHM) of the RSAM resonance is around $16\,$nm and the saturable
absorption recovery time is around $1\,$ps. The setup is shown in
Fig.~\ref{setup}. Both the VCSEL and RSAM are mounted on temperature
controlled substrates which allow for tuning the resonance frequency
of each cavity; parameters are set for having the emission of the
VCSEL resonant with the RSAM. The light emitted by the VCSEL is collected
by a large numerical aperture (0.68) aspheric lens and a similar lens
is placed in front of the RSAM. A $10\,\%$ reflection beam splitter
allows for light extraction from the external cavity and to monitor
both the VCSEL and the RSAM outputs. Intensity output is monitored
by a $33\,$GHz oscilloscope coupled with fast $10\,$GHz detector.
Part of the light is sent to two CCD cameras; the first one records
the near-field profile of the VCSEL, while the second records the
VCSEL's far-field profile. The external cavity length round-trip ($\tau$)
is fixed at $\tau=15\,$ns which corresponds to a $66.6\,$MHz fundamental
repetition rate.

\subsection{From conventional mode-locking to localized pulses}

The combination of the VCSEL and the RSAM in self-imaging condition
does not induce an appropriate ratio of the saturation parameters
for obtaining PML. Notwithstanding, mode-locking was obtained when
placing the RSAM surface in the exact Fourier transform plane of the
VCSEL near-field profile, i.e. when imaging the VCSEL far-field profile
onto the RSAM surface \cite{MJB-JSTQE-14}. As a consequence, the
VCSEL profile was imaged onto itself after a single external cavity
round-trip, but inverted (i.e. a magnification factor of $-1$). As
shown in \cite{MJB-JSTQE-14}, such configuration leads to the generation
of two opposed tilted plane waves, i.e. waves that have a propagation
wave-vector slightly out of the cavity axis $z$ and which travel
in the external cavity with an opposite transverse component and alternating
each other at every round-trip. Each one of these plane waves gives
birth to a train of mode-locked pulses separated by twice the external
cavity round-trip ($2\tau$), while the two trains are time shifted
of $\tau$. Intuitively, one understands that injecting the Fourier
transform of the VCSEL near-field profile into the RSAM strongly favors
the emission of a tilted wave from all points of the VCSEL, since
the Fourier transform of such an emission consists in a tight focused
single spot that saturates easily the RSAM. While this scheme leads
to conventional mode-locking and harmonic mode-locking for cavity
lengths and round-trips shorter than the gain recovery time ($\tau<\tau_{g}$),
we operate the system in the regime of long cavities, i.e. $\tau\gg\tau_{g}$,
and for bias currents below the laser threshold. In these conditions,
as shown in \cite{MJB-PRL-14}, several emission states, including
the zero intensity (off) solution, coexist for the same values of
bias current. 

\begin{figure}[tbh]
\centering{}\includegraphics[width=1\columnwidth]{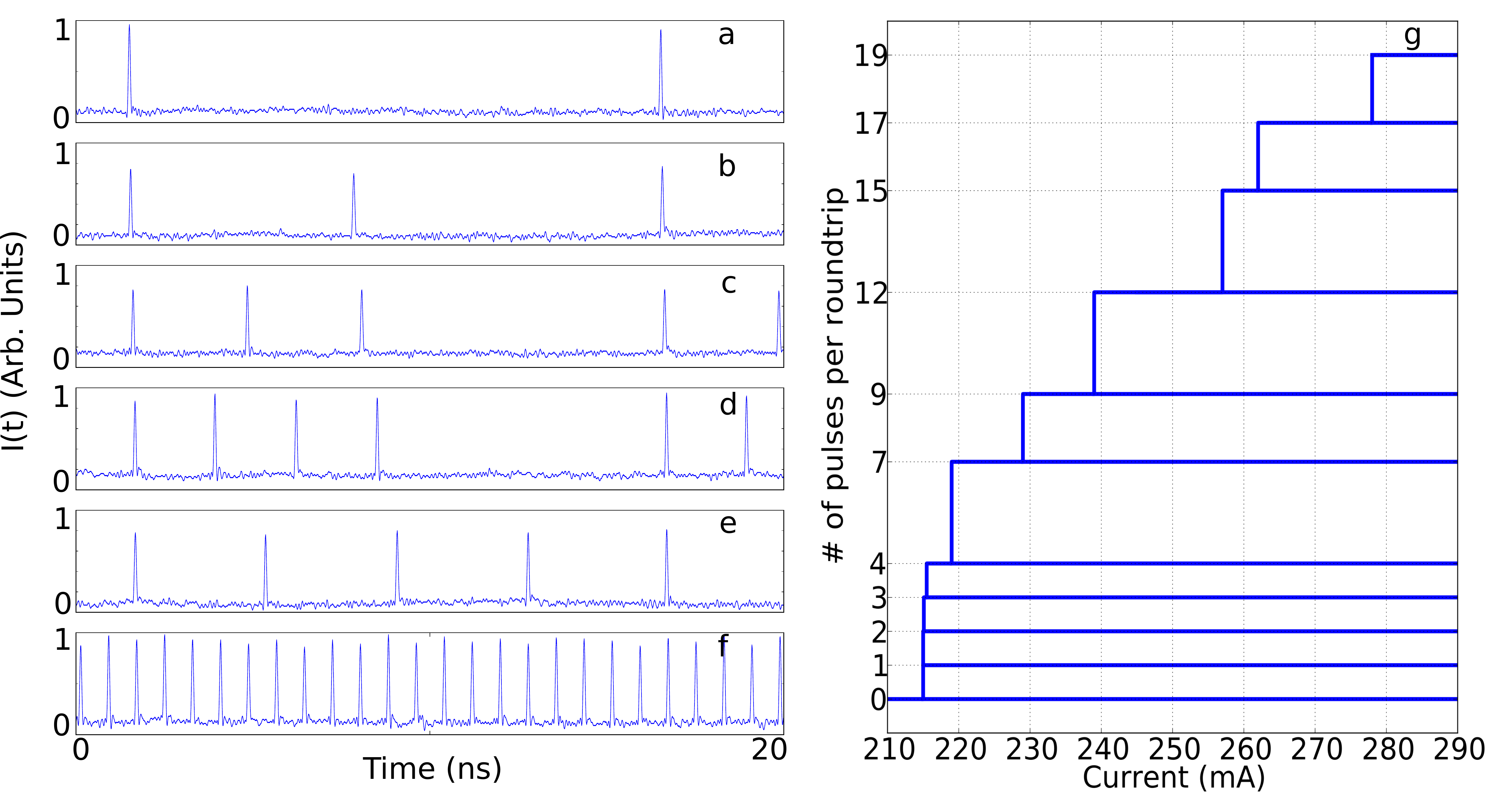}\caption{Left: Examples of coexisting time output traces for $J=$280~mA and
$\tau=15\,$ns. The panels a), b), c) and d) correspond to $N=1,2,3,4$
pulses per round-trip, respectively. Another configuration for $N=4$
is shown in e) where the pulses are regularly distributed within the
cavity. The maximal number of pulse is $N=19$ as depicted in panel
f). Right: experimentally obtained bifurcation diagram for the number
of pulses per round-trip (g) as a function of the bias current. The
stability of each solution is indicated by the solid horizontal lines.}
\label{traces}
\end{figure}

These emission states are characterized by a different number of pulses
circulating in the cavity, and for the same number of pulses, a large
variety of pulse arrangement can be observed. In Fig.~\ref{traces}a-d)
few examples between the large number of coexisting emission states
is represented, including $N=1,2,3$ and $4$ pulses circulating inside
the cavity. For the $N=4$ pulses case, we show two situations where
they appear either grouped, see Fig.~\ref{traces}d), or equally
separated Fig.~\ref{traces}e). The case of $N=19$ pulses circulating
within the cavity represents the largest number of pulses that can
be obtained for the size of the external cavity chosen. All the above
listed pulsing solutions coexist in the parameter space for a wide
range of VCSEL pumping current $J$. The pulse width cannot be determined
precisely from the oscilloscope traces shown in Fig.~\ref{traces},
which are limited by our real-time detection system (10~GHz effective
bandwidth). However, an estimation of the pulse width can be obtained
from the optical spectrum of the output, which exhibits a broad spectral
peak whose FWHM is around 0.12~nm that corresponds, assuming a time-bandwidth
product of 0.4, to a pulse width of 10~ps FWHM. The pulse was also
detected by a 36~GHz detector, which confirms a pulse width of less
than 12~ps FWHM considering the oscilloscope bandwidth limit. Finally,
the pulse peak power has been measured to be $1\,$Watt.

The multistability between a large number of different solutions is
shown in Fig.~\ref{traces}g), where we plot the bifurcation diagram
of these solutions as a function of the pumping current. Figure~\ref{traces}g)
is obtained increasing the parameter $J$ from $J=210\,$mA, where
only the steady off solution is stable, up to the value where this
solution loses its stability ($J<350\,$mA) and then sweeping it down
until a periodic emission with $N=19$ pulses per round-trip appears,
see Fig.~\ref{traces}f). In analogy to spatial localized structures
\cite{CRT-PRL-00,CoulletLSinfo}, this state corresponds to the fully
developed temporal pattern which is, together with the coexisting
stable off solution, at the origin of the localized structures. As
$J$ is decreased, the state with $N=19$ pulses per round-trip loses
its stability and the system bifurcates progressively towards states
with smaller numbers of LS, until a single one is present in the cavity.
Each state is spontaneously appearing as $J$ is scanned downward
and, once a new state appears, we increase $J$ to explore its stability
up to $J=290\,$ mA. As far as the system remains on the same branch
there is no change in the pulse arrangement, thus showing that, even
if several arrangements are possible, once one is chosen, it is stable
versus parameter changes. Figure~\ref{traces} indicates that in
this regime, mode-locked pulses in our system can be interpreted as
temporal localized structures, i.e. any pulse is independent from
the other and it can be individually addressed. It is worthwhile noting
that the localized character of the pulse implies that the pulse is
localized not only in the intensity but also in all the variables
describing the system. This means that, after the short intensity
pulse emission, the gain and saturable loss variables recover their
steady state value on a longer time scale thereby defining the effective
temporal extent of the LS. This is the main difference between localized
pulses and conventional mode-locked pulses. In the latter case, which
is obtained for pumping levels above laser threshold, the gain does
not recover to its steady state value between pulses. This is well
explained by the background stability criterion \cite{haus00rev}
which states that a pulsating solution is stable if losses are larger
the gain between pulses. Since the laser is pumped above threshold,
the only stable pulsating solution is the one having a number of pulses
in the cavity large enough such that, in-between pulses, gain cannot
recover sufficiently to overcome the unsaturated loss level. This
implies that fundamental mode-locking is possible only for $\tau<\tau_{g}$.
In the case of localized pulses, we are in the opposite limit and
the gain can recover fully between pulses because the maximal gain
level remains below the unsaturated loss level, even for an infinitely
long interval between pulses. Thus, localized PML pulses have no lower
limit in their repetition rate, as shown in Fig.~\ref{traces}a)
where a repetition rate of 66.6~MHz is obtained. On the other hand,
the recovery time of the slowest variable, in our case the gain recovery
time, determines the effective temporal extent of the localized pulse,
since no other pulse can occur before the recovery of all the variables.
Another important difference between localized pulses and conventional
mode-locked pulses is that, for a given set of parameters, the energy
of the localized pulses does not depend significantly on the number
of pulses circulating inside the cavity. Instead in conventional mode-locking
the energy of the system is shared between the pulses which circulate
inside the cavity and fundamental mode-locking is preferred over harmonic
mode-locking for maximum pulse energy.

\begin{figure}[b!]
\centering{}\includegraphics[width=1\columnwidth]{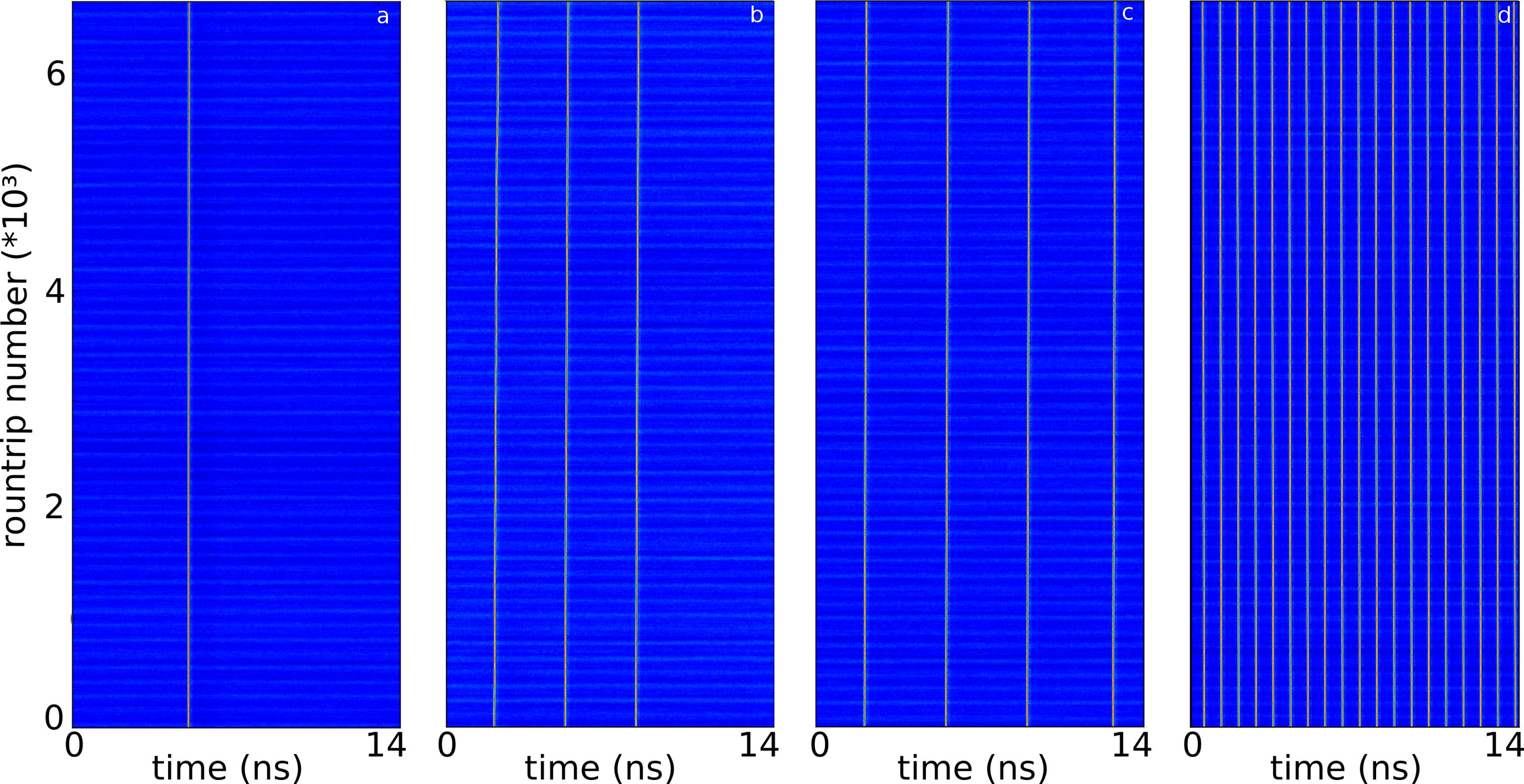}
\caption{Space-time like diagrams obtained. a) a single pulse per round-trip,
b) three pulses, c) four regularly spaced pulses and d) nineteen pulses
as in Fig.~\ref{traces} f).\label{diagcavlong} }
\end{figure}

In order to represent the evolution of the localized pulses travelling
within the cavity over many round-trips, we use the so-called space-time
diagram representation where the temporal trace is folded onto itself
at the cavity round-trip period. Accordingly, the round-trip number
$n$ becomes a pseudo-time variable while the pseudo-space variable
corresponds to the timing of the pulse modulo $\tau$. This representation
is similar to the one used for following the evolution of pulses along
optical fibers where fast and slow time scales are well separated
and it has been proposed also for delayed system \cite{AGL-PRA-92,GP-PRL-96,GJT-NAP-14}.
In Fig.~\ref{diagcavlong}a-d) we show some space-time like diagrams
obtained from the same time traces used for Fig.~\ref{traces} where
all the situations depicted coexist for the same parameter values.
Panel a) shows the evolution of a single LS in the external cavity,
while panel b), c) and d) represent the evolution of respectively
$N=3,\,4$ and $19$ pulses. In all the situations, the pulses do
not visibly jitter over the considered timescale of $6.5\times10^{3}\tau\sim=100\,\mu$s
and they seems to be insensitive to the surrounding noise. This can
be attributed to the presence of the saturable absorber which acts
as an effective noise eater. While the RSAM opens a very short window
of net gain around the pulse, the remaining low intensity emission
from the VCSEL is absorbed by the RSAM. Accordingly, the background
shown in Fig.\ref{diagcavlong} is homogeneous and corresponds to
zero emission. 

In presence of localized structures, as the ones shown in Fig.~\ref{diagcavlong},
the possibility of forming bounded states of localized structures,
also called molecules, naturally arises.When gathered in molecules,
the pulses are not independent anymore, while the molecule itself
becomes a compound localized structure which is independent from other
localized structures that may travelling within the cavity. Such molecules
have been widely observed in fiber lasers \cite{GA-NAP-12} and in
general only a few if not an unique equilibrium distance separates
the ``atoms'' of a molecule. The signature of such bound states
of LS in a space-time like diagram is the presence of preferred distances
between two neighbor pulses. We have never found any preferred distances
between close pulses in the space-time like diagrams that we have
analyzed. Instead we have observed a \emph{continuum} of temporal
separations. One can understand the absence of molecules in our system
by the specificity of the interaction between neighbor LS. Indeed,
it is known that the gain depletion induced by each LS produces a
repulsive interaction \cite{TK-JQE-09} that decays over a time scale
$\tau_{g}$. The potentially attractive interaction mediated by the
oscillating tails of the pulses vanishes after a few times the effective
pulse width, i.e. a time scale of $\sim30\,$ps. As such, this clear
scale separation between the attractive and repulsive forces explain
why LS molecules would be hard to obtain with semiconductor active
materials. 

It may be useful to compare our results with the extensive literature
on mode-locking. Localized pulses in mode locked laser might have
been observed in the past. For example, several papers mention PML
regimes obtained at \emph{exceedingly low} frequencies with respect
to the limits of the background stability criterion; repetition rates
of the order of few hundreds MHz were reported in\cite{MTP-JSTQE-11,CYN-JSTQE-11,BSK-EL-13}.
We believe that these regimes could be explained in terms of LS. While
it is difficult to know exactly the experimental conditions of these
works, in \cite{MTP-JSTQE-11} it is clearly mentioned that the off
solution coexists with the PML solution. The experimental result reported
recently in \cite{BVE-OE-13}, claiming mode-locking at a repetition
rate of 85.7~MHz, can also be explained in terms of LS. In this work,
neighboring pulses of $50\,$ps FWHM, separated of $1\,$ns, are interpreted
theoretically as bound states. While the experimentally observed distance
corresponds to the vanishing of the repulsive interaction induced
by the gain depletion \cite{TK-JQE-09}, the theoretical analysis
is performed after adiabatically eliminating the carrier dynamics,
thus neglecting such repulsive interactions. We believe that this
approximation in \cite{BVE-OE-13}, correct for fiber lasers but not
for semiconductor devices, has led to a wrong interpretation of independent
neighbor pulses in terms of bound states resulting of the locking
of the oscillating tails of the pulses.

\subsection{Localized pulse addressing}

\begin{figure}[tbh]
\begin{centering}
\includegraphics[bb=0bp 0bp 1603bp 738bp,clip,width=0.5\textwidth]{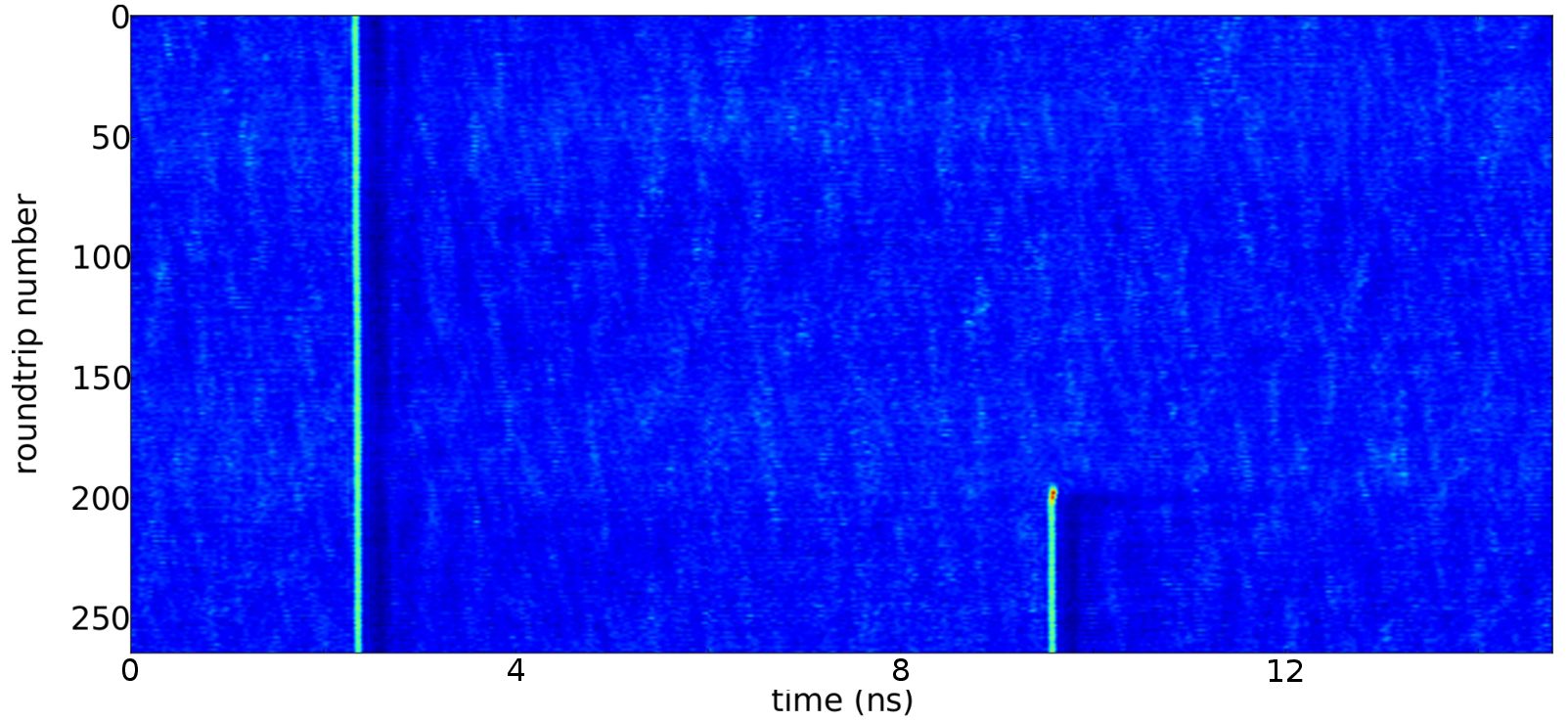}
\par\end{centering}

\caption{Space-time like diagram showing nucleation of a localized pulse after
a mechanical perturbation of the system. Parameters are in the multi-stable
region of Fig.~\ref{traces} }
\label{allumage}
\end{figure}

The most important property of localized structures is their mutual
independence which allows for their individual addressing. The localized
pulses that we have obtained can be individually triggered by shining
light pulses inside the cavity, as we have shown numerically in \cite{MJB-PRL-14}.
By encoding an information bit in the form of a localized pulse, our
system can be operated as an all-optical buffer, similar to the one
proposed using fiber Kerr resonator with a driving field \cite{LCK-NAP-10},
but taking advantage of the compactness and fast response of semiconductor
materials as in \cite{GJT-NAP-14}. The buffer bit-rate depends on
the minimal time interval between two independent pulses, i.e. the
temporal extent of the LS which is, in our case, fixed by the gain
recovery time. The experimental results shown in Fig. \ref{traces}f)
indicate that a separation of 1~ns is certainly sufficient to avoid
interaction between pulses, thus leading to buffer bit rate of $\sim1\,$GHz.
While the all-optical addressing of the mode-lock localized pulses
is the topic of ongoing research, the nucleation of a single pulse
can be observed when perturbing mechanically the system, as shown
in Fig.~\ref{allumage}. Yet this kind of perturbation does not allow
any control of the switching; it may lead to the nucleation of several
pulses at the same time with random positions within the cavity. 

\begin{figure}[h]
\centering{}\includegraphics[bb=0bp 0bp 1450bp 783bp,clip,width=1\columnwidth]{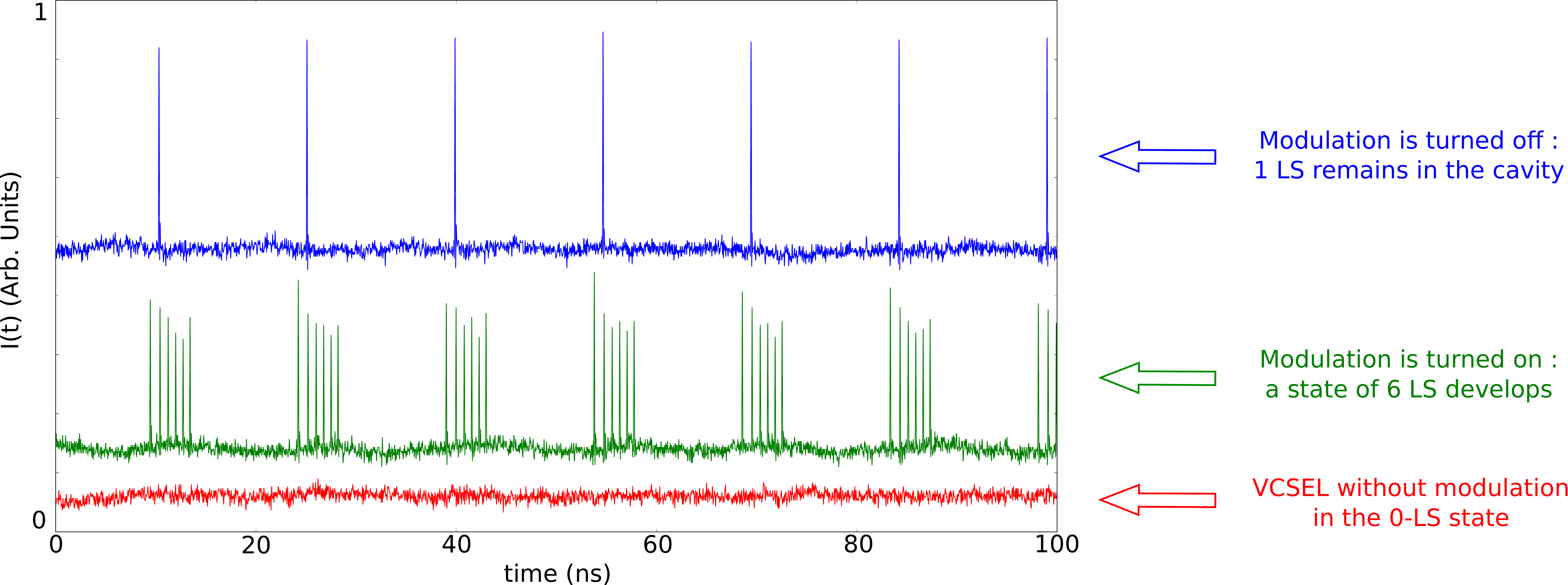}
\caption{Time traces showing the experimental process for LS generation via
current modulation. First the laser is in the 0-LS state without modulation
(red, bottom line) and $J=215\,$mA. Second, modulation is applied
with a peak-to-peak amplitude of 460~mA and 6~LS appear close to
the peak of the modulation (green, middle line). When we switch-off
current modulation, only a single LS remains in the cavity (blue,
top line). \label{buffertrace} }
\end{figure}

Notwithstanding, a higher degree of control can be implemented when
perturbing the bias current of the VCSEL. The main limitation using
this parameter comes from the frequency response of the VCSEL to a
modulated signal coupled onto the pumping current via a bias T. The
electrical characteristics of the broad-area VCSEL we used, together
with the bandwidth of the bias-T used, limit the frequency modulation
below 0.3~GHz. Accordingly, the individual addressing of localized
pulses via the emission of electrical pulses suffers from the too
long rise time of the perturbation. Nevertheless a sinusoidal modulation
of the pumping current with a period corresponding to the cavity round-trip
$\tau$ allows for the excitation of localized pulses in a narrow
time interval within the round-trip time, i.e. a well-defined vertical
band of the space-like variable, as plotted in Fig.~\ref{diagcavlong}.
The idea would be to modulate the current between a lowest value where
the off solution is the only possible solution up to a highest value
where the fully developed temporal pattern shown in Fig.~\ref{traces}f)
is also the unique possible situation. Since this modulation is synchronous
with the cavity round-trip, a portion of this monostable temporal
pattern will appear only in correspondence with the modulation signal
peak. Once the sinusoidal perturbation of the pumping current is removed,
the portion of the monostable pattern will give way to a solution
with one, two or more localized pulse depending on the continuous
bias current value of the VCSEL, see Fig. \ref{traces}g), and on
the size of the the temporal pattern excited. The result of this experiment
is shown in Fig.~\ref{buffertrace} where the VCSEL is biased at
$J=215\,$mA, a current value for which for the sake of simplicity
only the state with a single localized pulse per round-trip coexists
with the off state. We prepare the system in the off state and then
we apply a modulation having a peak to peak amplitude of 460~mA.
In correspondence with the modulation peak (where $J=445\,$mA), six
localized pulses appear at a distance of 800~ps, which corresponds
to the interpulse distance of Fig.~\ref{traces}f), i.e $\sim\tau/19$.
Once the modulation signal is removed, the continuous value of pumping
current does not support the stable existence of six localized pulses
and the system bifurcates towards the state with a single pulse per
period. This regime is stable and it persists indefinitely.

\subsection{Multi-Pulse pattern generation}

The experiment reported above suggests that a current modulation can
be used to obtain a pattern of closely packed localized pulses. This
functionality is appealing for many applications like e.g. OCDMA \cite{ocdma}
and LIDAR \cite{lidar,lidar2}, since narrow optical pulses (<10 ps)
are typically obtained using conventional PML systems which deliver
a periodic train of pulses with no versatility in the pulse arrangement.
Pulse patterning can be of course obtained by adding optical gates
based on Electro-Optical Modulators, but these solutions are expensive,
energetically inefficient, and their implementation requires synchronization
between the pulse source and the gating. We implement a pulse pattern
generator by driving the system pumping current with a forcing at
a period equal to the cavity round-trip and which brings the system
into the monostable region where only the periodic pattern of pulses
is stable, according to Fig.~\ref{traces}f). The amplitude of the
forcing together with the steady value of the bias current, fix the
time interval during which the system is driven in this monostable
region and, accordingly, this fixes the number of pulses which are
excited. For a stable output, the modulation period needs to match
closely the cavity round-trip $\tau$, which in our case requires
a modulation frequency of $\nu_{mod}$ = 66.584.030~Hz. We have verified
that a stable output pattern is lost if the frequency mismatch exceeds
$10\,$KHz.

\begin{figure}[b!]
\begin{centering}
\includegraphics[width=0.95\columnwidth]{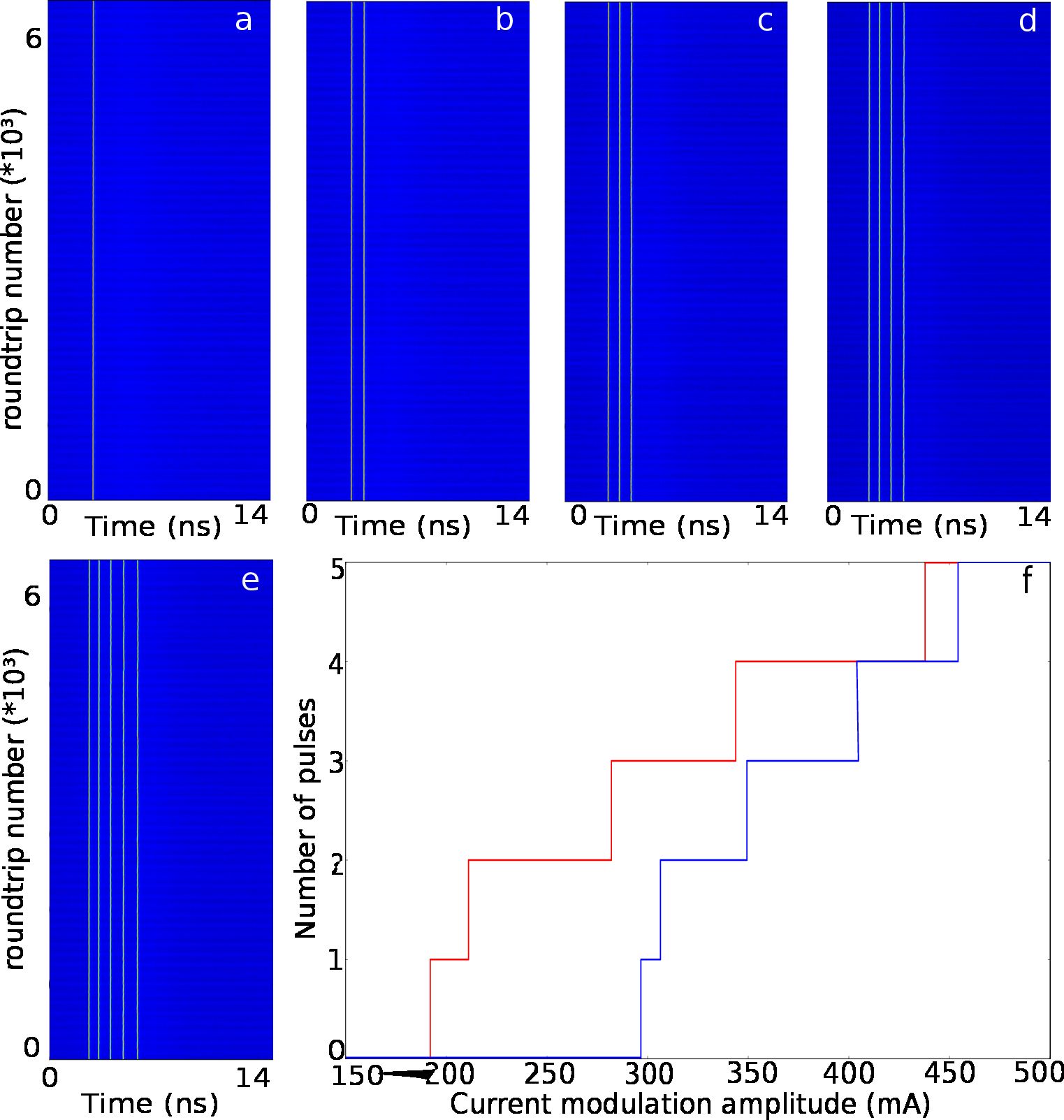} 
\par\end{centering}

\caption{Modulation of the bias current can be used to address temporal LS
in the cavity. $J=210\,$mA, $\tau=14.8\,$ns, $\nu_{mod}$ = 1/$\tau$.
As the modulation peak-to-peak amplitude is increased a) 297~mA,
b) 306~mA, c) 349~mA, d) 405~mA and e) 454~mA, the number of pulses
close to the peak of the modulation is incremented by one. f) The
reverse sequence appears when decreasing the modulation amplitude.
The modulation value at which transition from $N-1$ pulses to $N$
pulses happens is slightly different with respect the modulation value
at which the inverse transition from $N$ pulses to $N-1$ occurs,
denoting a strong multistability, . \label{address} }
\end{figure}

The results obtained are shown in Fig.~\ref{address} where we control
the number of pulse emitted from one to five by changing the amplitude
of the sinusoidal forcing. The distance between pulses is fixed by
the effective temporal extent of the pulses which, as explained above,
depends on the carrier recombination time. The close inspection of
panel e) and f) reveals that the distance between pulses can change
slightly within a pattern. This can be explained by the use of a sinusoidal
signal for bringing the system in the monostable region of Fig.~\ref{traces}f),
where a periodic pattern of pulses is emitted. In the portion of the
sinusoidal signal where this occurs, the current is still increasing,
thus leading to an effectively faster carrier recovery rate and therefore
to closer pulses. For the same reason, pulse intensities vary slightly
inside the pattern. The appearance of an increasing number of pulses
for larger amplitudes of the forcing, as well as the inversed phenomenon
while decreasing the modulation strength, occurs with an high degree
of hysteresis as seen in Fig.~\ref{address}f), thus revealing multistability
between the different situations depicted in Fig.~\ref{address}a)-e) 

\begin{figure}[h]
\centering{}\includegraphics[width=0.95\columnwidth]{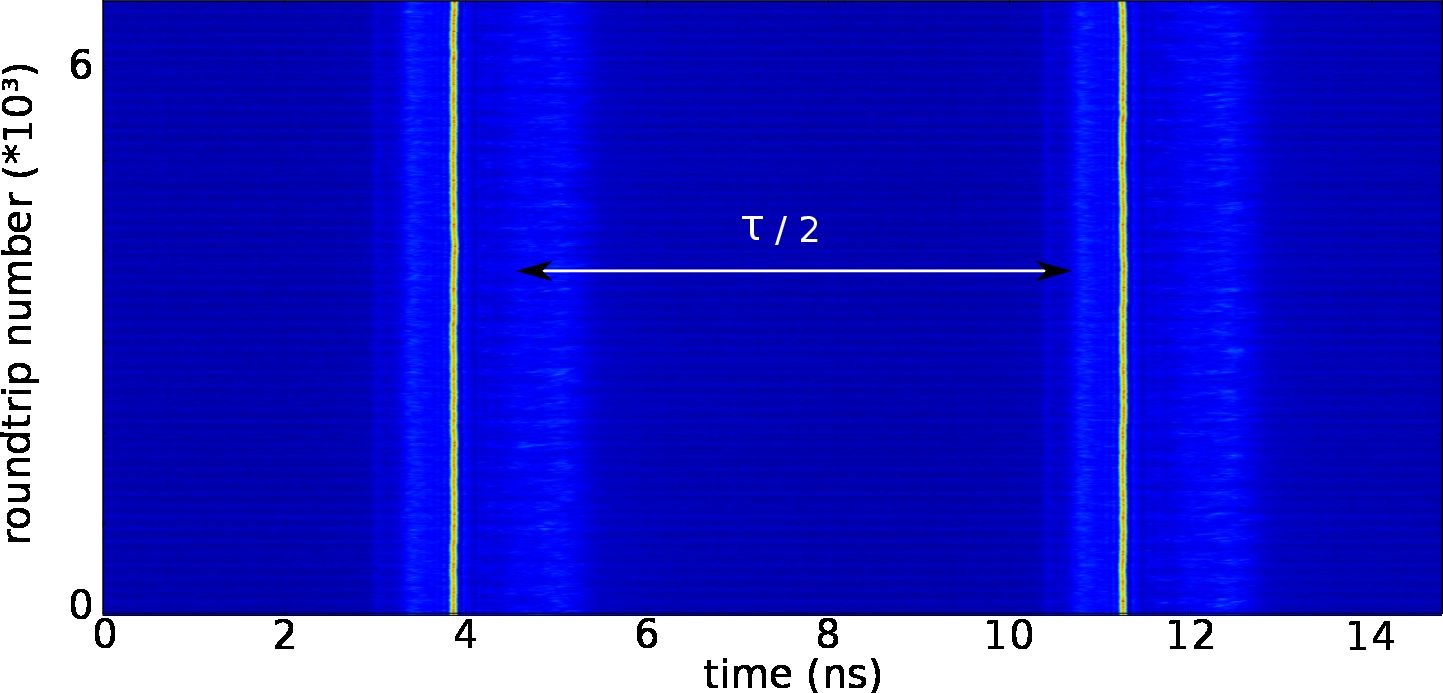}
\caption{Writing the LS is possible even when the modulation period is set
to $\tau/2$. The bias is $J=210\,$mA, and the modulation amplitude
$297\,$mA.\label{address2FSR} }
\end{figure}

While Fig.~\ref{address} discloses the proof-of-principle of a pulse
pattern generator based on localized pulses, more complicated pulse
pattern structures can be envisioned using other kind of modulation
signals, as square or pulsed current modulation. Even if these modes
of operation require to solve the problem of the electrical coupling
with the VCSEL, we have tried to use a sinusoidal signal at a period
corresponding to $\tau/2$ for addressing two patterns of pulses within
the cavity. The result is shown in Fig.~\ref{address2FSR}, where
two localized pulses are separated by half of the round-trip time,
in correspondence to the modulation peaks.

\section{Theoretical results}

The proper study and the simulation of PML lasers is a demanding problem
from the computational point of view: while pulses may form on a relatively
short time scale of a few tens of round-trips, the pulse characteristics
only settle on a much longer time scale \cite{JB-JQE-10}. If anything,
the complex transverse dynamics \cite{MJB-JSTQE-14} also present
in our case shall slow down the dynamics even further. In particular,
we consider here extremely long cavities and therefore very long delays
for which with the pulse to round-trip aspect ratio is $\sim10^{3}$.
The necessary presence of noise prevents the use of adaptive step-size
algorithms, which would be particularly suitable for such strongly
nonlinear problems.

Hence, for the sake of simplicity, we use a purely temporal model
as for instance the generic delay differential equation model of \cite{VT-PRA-05}
which generalizes Haus' model as it encompasses both the pulsating
and the steady regimes. While more detailed results could also be
obtained with a traveling-Wave approach \cite{Freetwm}, we found
that such a simple model was sufficient to give a good qualitative
agreement with the experimental results. Denoting by $A$ the amplitude
of the optical field, $G$ the gain, and $Q$ the saturable absorber
losses, the model reads
\begin{eqnarray}
\frac{\dot{A}}{\gamma} & \negthickspace=\negthickspace\negthickspace & \sqrt{\kappa}\exp\left[\frac{\left(1-i\alpha\right)G_{\tau}-\left(1-i\beta\right)Q_{\tau}}{2}\right]A_{\tau}-A,\label{eq:VT1}\\
\dot{G} & \negthickspace=\negthickspace\negthickspace & g_{0}+\Delta g\sin\left(\frac{2\pi t}{\tau_{m}}\right)-\Gamma G-e^{-Q}\left(e^{G}-1\right)\left|A\right|^{2},\label{eq:VT2}\\
\dot{Q} & \negthickspace=\negthickspace\negthickspace & q_{0}-Q-s\left(1-e^{-Q}\right)\left|A\right|^{2},\label{eq:VT3}
\end{eqnarray}
where time has been normalized to the SA recovery time, $\alpha$
and $\beta$ are the linewidth enhancement factors of the gain and
absorber sections respectively, $\kappa$ is the fraction of the power
remaining in the cavity after each round-trip, $g_{0}$ is the pumping
rate, $\Delta g$ and $\tau_{m}$ the amplitude and the period of
the modulation of the gain, $\Gamma$ is the gain recovery rate, $q_{0}$
is the value of the unsaturated losses which determines the modulation
depth of the SA, $s$ is the ratio of the saturation energy of the
SA and of the gain sections and $\gamma$ is the bandwidth of the
spectral filter. In Eq.~\ref{eq:VT1} the subscript $\tau$ denotes
a delayed value of the variable, $x_{\tau}=x\left(t-\tau\right)$.
This delay renders the dynamical system infinitely-dimensional and
it describes the spatial boundary conditions of a cavity closing onto
itself. As such it governs the fundamental repetition rate of the
PML laser. We use standard parameter values that closely represents
the experimental situation: $\kappa=0.8$ , $s=15$, $q_{0}=0.3$,
$\Gamma^{-}$$^{1}=66.66$, $\gamma=3$ and $\tau=3000$. Assuming
a SA recovery time of $5\,$ps, this corresponds to a gain recovery
of $333\,$ps, a bandwith FWHM of $400\,$GHz, i.e. $1.3\,$nm around
$980\,$nm in good agreement with the VCSEL resonance linewidth measurement
with a tunable laser and a time delay of $15\,$ns. Importantly, we
stress that the existence of PML regimes below the lasing threshold
is mostly independent of the phase-amplitude couplings which explains
while in the past we choose $\alpha=\beta=0$ for the sake of simplicity.
However, we set here the more realistic values of $\alpha=1.5$ and
$\beta=1$. The lasing threshold is determined by the pump level $g_{0}$
for which the off solution $\left(A,G,Q\right)=\left(0,\Gamma^{-1}g_{0},q_{0}\right)$
becomes linearly unstable; in our case it is given by $g_{th}=\Gamma\left(q_{0}-\ln\kappa\right)$.
Above threshold, the continuous wave (CW) solution bifurcates supercritically
from the off state, but PML is found below threshold where the pulsating
branches emerge as saddle-node bifurcation of limit cycles, see \cite{MJB-PRL-14}
for more details.

\subsection{Multi-pulse pattern}

\begin{figure}[h!]
\centering{}\includegraphics[bb=0bp 0bp 440bp 320bp,width=1\columnwidth]{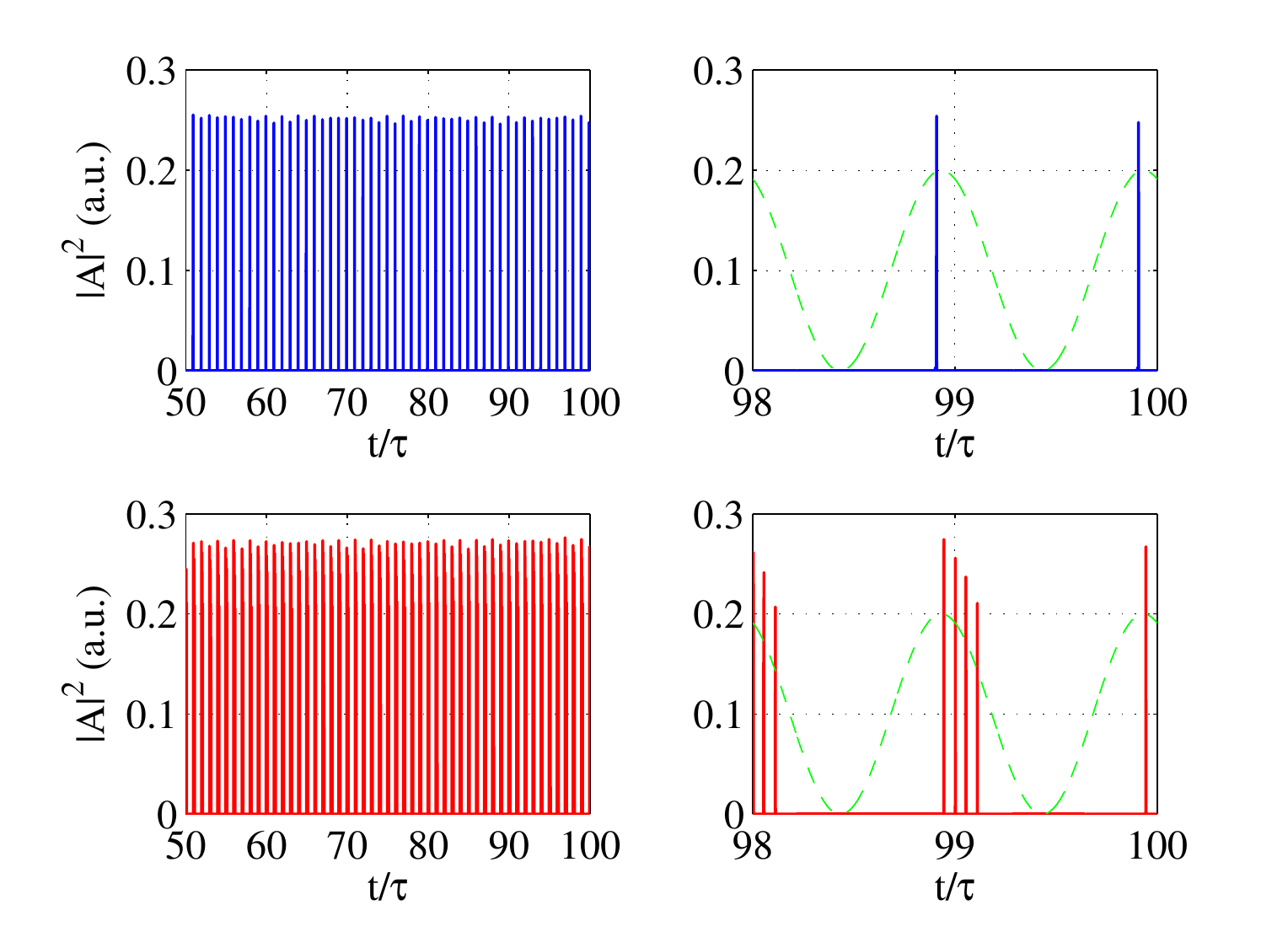}
\caption{Different co-existing time traces of the temporal intensity over several
and two round-trips for $\Delta g=0.23$ and $\tau_{m}=\tau+0.2$.
The dashed green lines (not to scale) represent the modulation of
the gain. \label{theo_modJ_trace}}
\end{figure}

We set the modulation period very close to the delay value as $\tau_{m}=\tau+0.26$,
the choice of such a precise value will become clear in a moment.
The DC bias current is below threshold $g_{0}=0.8g_{th}$ but also
below the minimal current for which LS exists that we denote $g_{sn}\sim0.93g_{th}$.
As such, we are in a situation similar to the experiment. We depict
our results in Fig.~\ref{theo_modJ_trace} where we evidence the
existence of bistability between different temporal patterns composed
of either one or three pulses. One clearly see that the pulses organize
in the region where the gain is maximal. An inspection of the gain
and of the saturable absorption helps to clarify the mechanism of
formation of these pulses. During a fraction of the round-trip, the
gain is above $g_{sn}$ while it remains still below $g_{th}$. As
such one or several pulses can be generated. The pulses do interact
via the gain dynamics which is apparent in the fact that they do not
have exactly the same heights. Gain dynamics is known to induce repulsive
interaction \cite{TK-JQE-09} between pulses. Intuitively, pulses
will move toward higher gain, which corresponds to maximizing the
distance between them, hence explaining the repulsive nature of the
interaction. However, here they cannot move away from each other since
the gain varies in time an go below the minimal value at which they
can exists. As such, the equilibrium distance between pulses comes
for the complex interplay between the repulsive interaction between
pulses and the presence of the modulation that acts as an external
potential. 

\begin{figure}[h!]
\centering{}\includegraphics[width=1\columnwidth]{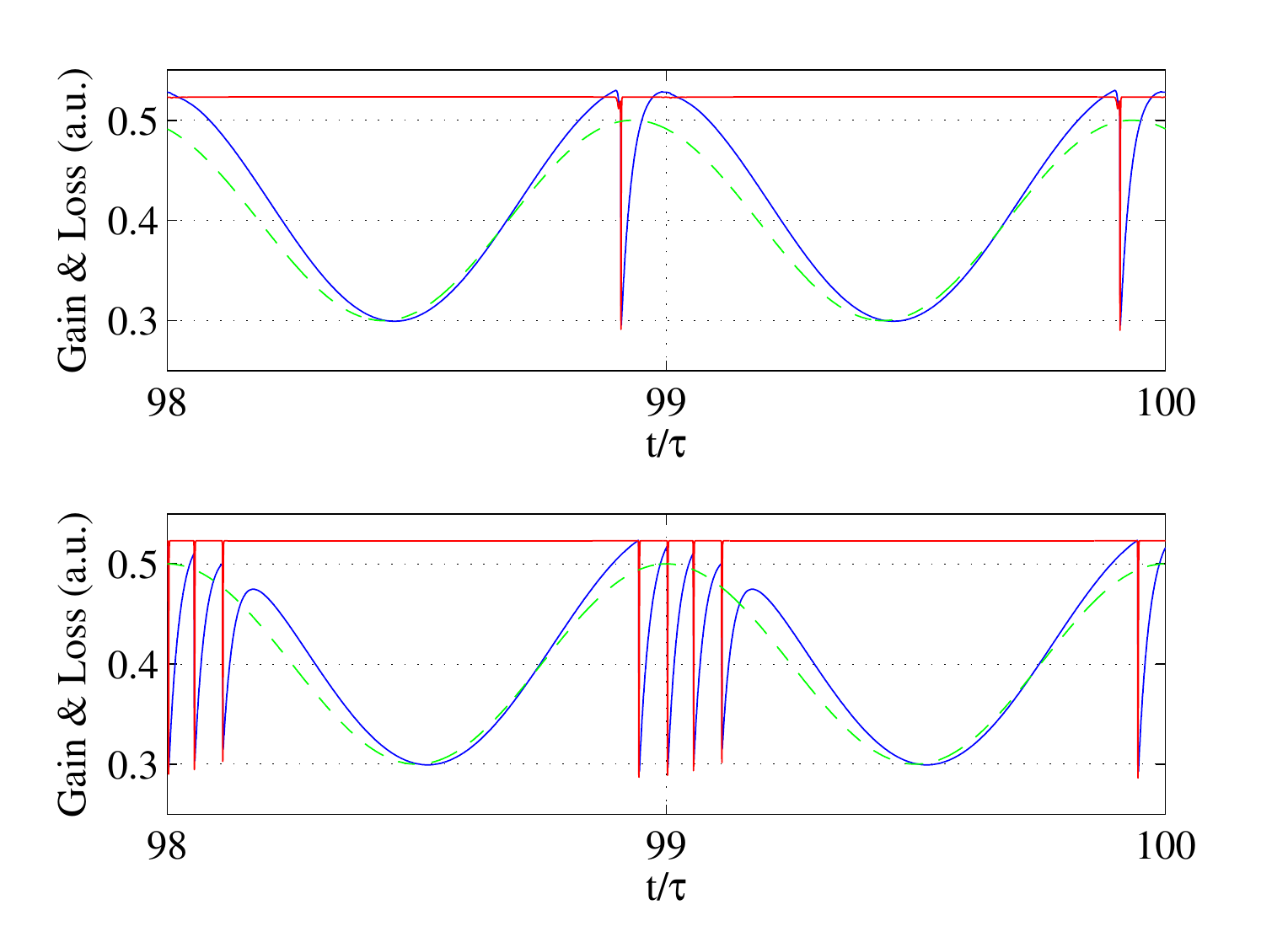}
\caption{Gain and absorption dynamics for the two co-existing time traces in
Fig. \ref{theo_modJ_trace} over two round-trips with $\Delta g=0.23$
and $\tau_{m}=\tau+0.2$. The dashed green lines (not to scale) represent
the modulation of the gain. \label{theo_modJ_GQ}}
\end{figure}

\begin{figure}[h!]
\centering{}\includegraphics[bb=25bp 0bp 410bp 160bp,clip,width=1\columnwidth]{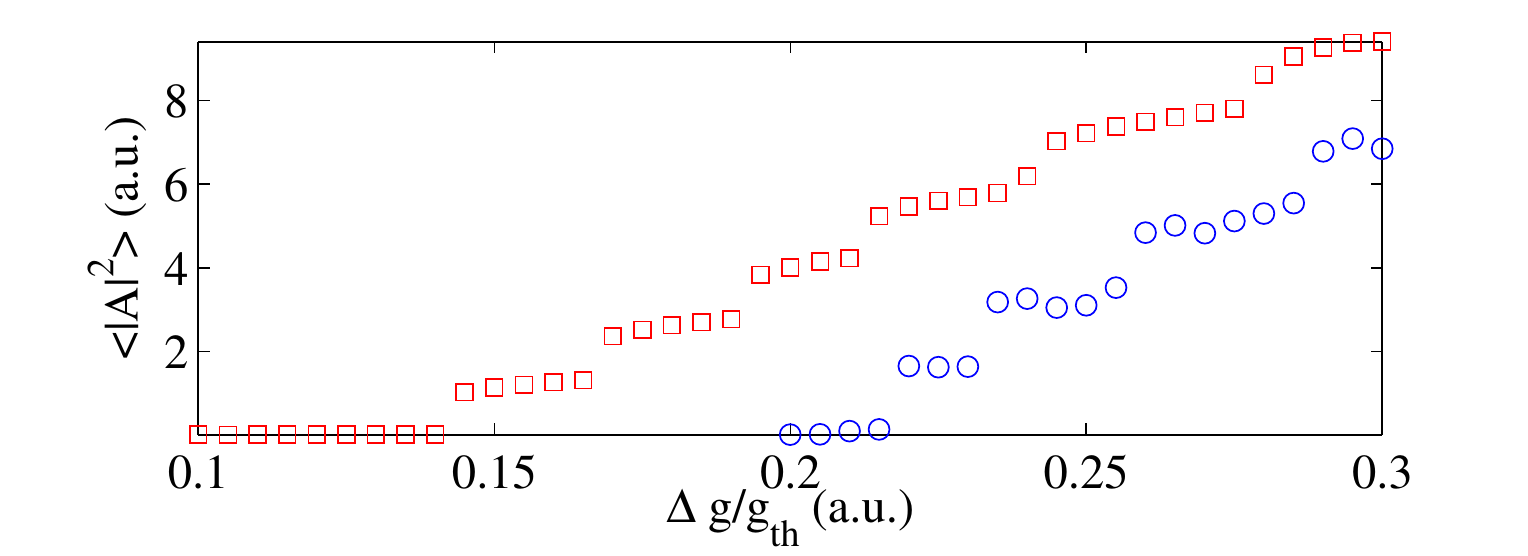}
\caption{Numerical bifurcation diagram for the pulse number as a function of
the modulation of the gain setting $g_{0}=0.8g_{th}$. We represent
the (normalized) average intensity over a round-trip. The blue circles
(resp. red squares) correspond to an upward (resp. downward) scan
of the modulation amplitude signaling a strong multi-stability. Each
jump corresponds to the appearance of the disappearance of an additional
LS. \label{theo_modJ_diag}}
\end{figure}

A rich multistability diagram for the number of pulse as a function
of the gain modulation $\Delta g$ can be appreciated in Fig.~\ref{theo_modJ_diag},
in agreement with experimental results of Fig.~\ref{address}f).
For increasing modulation starting from the off solution, we find
in Fig.~\ref{theo_modJ_diag} that small amplitude broad peaks appear
for $\Delta g=0.2g_{th}$. This corresponds to the laser reaching
exactly threshold at some instant since $g_{0}=0.8g_{th}$ and such
low and broad intensity pulses (not shown) can be interpreted as locally
amplified spontaneous emission. The first real LS appears at $\Delta g=0.23g_{th}$.
Upon increasing further the modulation amplitude, the pulses energy
slowly increases which explains the slope of the plateau (the experiment
considering only the pulse number) while the jumps correspond to the
apparition of another pulse. Notice in addition that the pulses within
a group may have slightly different intensities. Such an information
is not apparent in Fig.~\ref{theo_modJ_diag} as we compute the average
intensity over a round-trip. Upon decreasing the bias, a strong hysteresis
is apparent as in Fig.~\ref{address}f). Interestingly, one is able
to reach much lower values for the modulation amplitude. The value
of $\Delta g=0.14g_{th}$ corresponds to a maximal gain of $0.94g_{th}$
which is very close to the value of $g_{sn}$, i.e. the minimal value
of the current in DC allowing for the existence of a single LS. The
diagram in Fig.~\ref{theo_modJ_diag} is normalized in such a way
that the integrated intensity of a single pulse at the minimal bias
current is unity. We remark that if we have bistability between 1
and 4 pulses, say for instance at $\Delta g=0.22g_{th}$, this certainly
implies that we can also have 2 and 3 pulses. As such the diagram
of Fig.~\ref{theo_modJ_diag} only gives the extreme envelope of
a more complex multi-stable diagram.

\begin{figure}[h!]
\centering{}\includegraphics[bb=0bp 0bp 360bp 160bp,width=1\columnwidth]{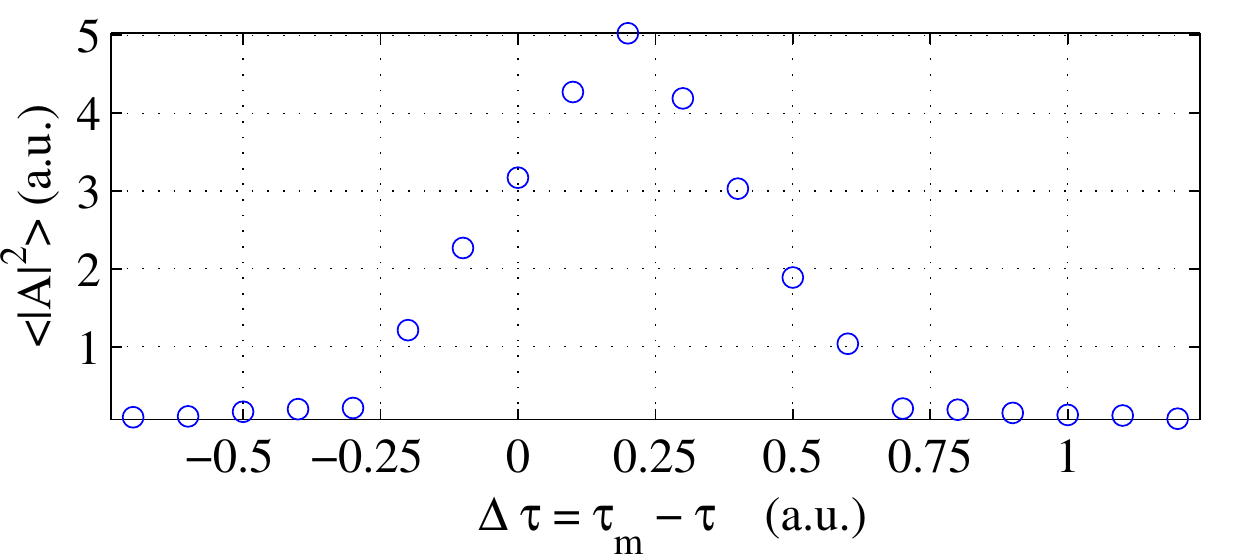}
\caption{Numerical bifurcation diagram for the pulse number as a function of
the detuning of the modulation period as compared to the cavity round-trip.
Once notice that a maximal number of pulse is obtained for $\Delta\tau=0.23$
which corresponds indeed to the effective round-trip in the cavity
dressed by the active medium response time. One notice the sharpness
of such a resonance tongue whose width is typically proportional to
the pulsewidth. \label{theo_modJ_scantau}}
\end{figure}

Experimental results also shown a strong sensitivity of dynamics with
respect to the modulation period. We notice that the period of the
PML regimes scales as $T\sim\tau+\gamma^{-1}$, see \cite{VT-PRA-05}
for instance. The pulsewidth is proportional to the inverse of the
filter bandwidth and there is a deep relation between pulsewidth and
deviation of the pulse train period from the value given by the time
of flight. One can intuitively consider that $\gamma$ is the ``inertia''
of the filter. As such, if a short pulse, say a Dirac delta, is re-injected
after a time of flight $\tau$, the filter needs a typical time $\gamma^{-1}$
to filter and re-emit it. In our case, we found that for $g_{0}=0.8g_{th}$
the fundamental period is $\tau_{m}=\tau+0.26$ which explains why
we choose this value in our analysis. We depict in Fig.~\ref{theo_modJ_scantau}
the sensitivity to the modulation period for a low amplitude modulation
of $\Delta g=0.23g_{th}$. One can verify from Fig.~\ref{theo_modJ_diag}
that in this case we can have either 1 or 4 pulses. In order to have
a simple picture, we start from an initial condition with as many
pulses as possible, i.e. the harmonic mode-locking of maximal order
found just threshold at $g_{0}=1.2g_{th}$ in the absence of modulation
$\Delta g=0$. Then we suddenly decrease $g_{0}$ toward $g_{0}=0.8g_{th}$
and set the modulation to $\Delta g=0.23g_{th}$. Such a method allows
us to find the maximal number of packed pulse that the system can
support as a function of the detuning of the modulation frequency.
We see in Fig.~\ref{theo_modJ_scantau} that indeed a precise tuning
is necessary as the full width of the resonance tongue is $0.5$,
i.e. of the same order of magnitude than the pulsewidth. Saying that
the resonance tongue if of the order of the pulsewidth yields a bandwidth
of $\Delta\nu\sim10\,$ps/$\tau^{2}\sim40\,$kHz in good qualitative
agreement with the experimental results.

\section{Conclusions}

We have shown that electrically biased broad-area VCSELs with optical
feedback from a RSAM can be operated in a regime where the passively
mode-locked pulses can be addressed and controlled individually when
the compound system is operated below threshold. The strong multistability
between the off solution and a large variety of pulsating solutions
with different number and arrangements of pulses per round-trip, demonstrate
that the mode-locked pulses are mutually independent. We show how
a modulation of the bias current allows controlling the number of
the pulses travelling within the cavity, paving the way an arbitrary
pattern generator of picosecond pulses.

\section*{Acknowledgment}

J.J. acknowledges financial support from the Ramon y Cajal fellowship
and the CNRS for supporting a visit at the INLN where part of his
work was developed as well as useful discussions with H. Wenzel. J.J.
and S.B. acknowledge financial support from project RANGER (TEC2012-38864-C03-01)
and from the Direcció General de Recerca, Desenvolupament Tecnològic
i Innovació de la Conselleria d\textquoteright{}Innovació, Interior
i Justícia del Govern de les Illes Balears co-funded by the European
Union FEDER funds. M.M. and M.G. acknowledge funding of Région PACA
with the Projet Volet Général 2011 GEDEPULSE and ANR project OPTIROC.


\def\url#1{}

\begin{IEEEbiography}[{\includegraphics[width=2.5cm]{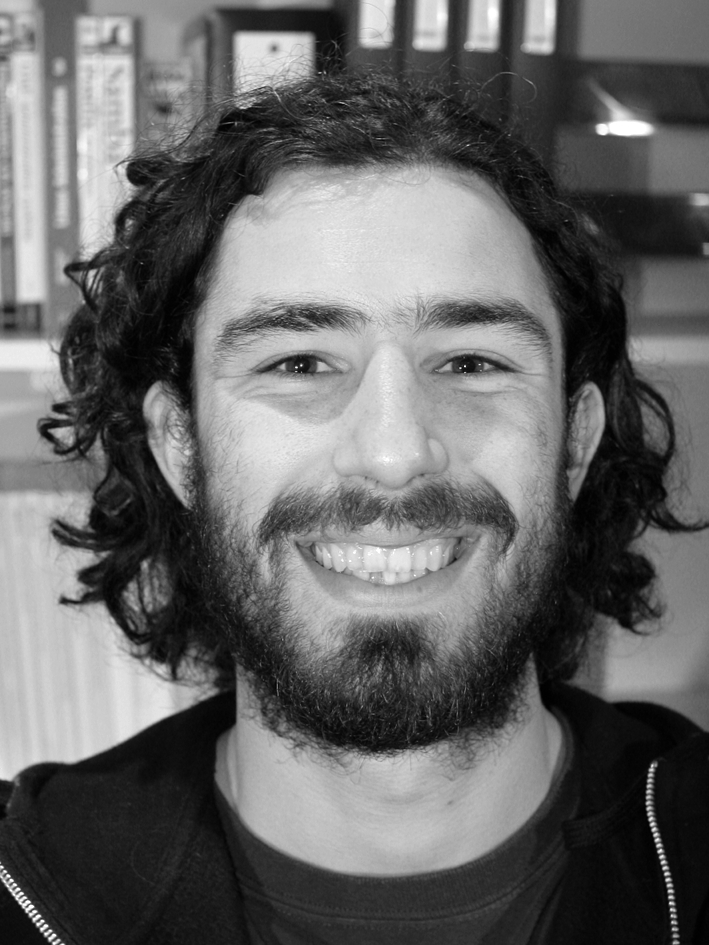}}]
{Mathias Marconi} was born in Nice, France, in 1988. From 2006 to
2011, he was a student at Université de Nice Sophia Antipolis (UNS),
France. In 2011, he was an exchange student at Strathclyde university,
Glasgow, U.K. The same year he obtained the Master degree in optics
from UNS. He obtained the Ph.D. degree at the Institut Non-linéaire
de Nice, Valbonne, France in 2014. His research interests include
semiconductor laser dynamics and pattern formation in out of equilibrium
systems. He is a member of the European Physical Society. 
\end{IEEEbiography}

\begin{IEEEbiography}[{\includegraphics[width=2.5cm]{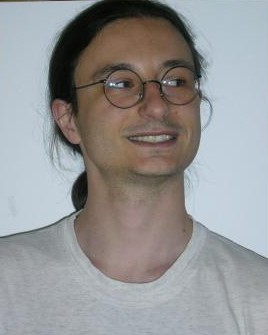}}]
{Julien Javaloyes} (M\textquoteright{}11) was born in Antibes, France
in 1977. He obtained his M.Sc. in Physics at the ENS Lyon the PhD
in Physics at the Institut Non Linéaire de Nice / Université de Nice
Sophia-Antipolis working on recoil induced instabilities and self-organization
processes in cold atoms. He worked on delay induced dynamics in coupled
semiconductor lasers, VCSEL polarization dynamics and monolithic mode-locked
semiconductor lasers. He joined in 2010 the Physics Department of
the Universitat de les Illes Balears as a Ramón y Cajal fellow. His
research interests include laser dynamics and bifurcation analysis.
\end{IEEEbiography}

\begin{IEEEbiography}[{\includegraphics[width=2.5cm]{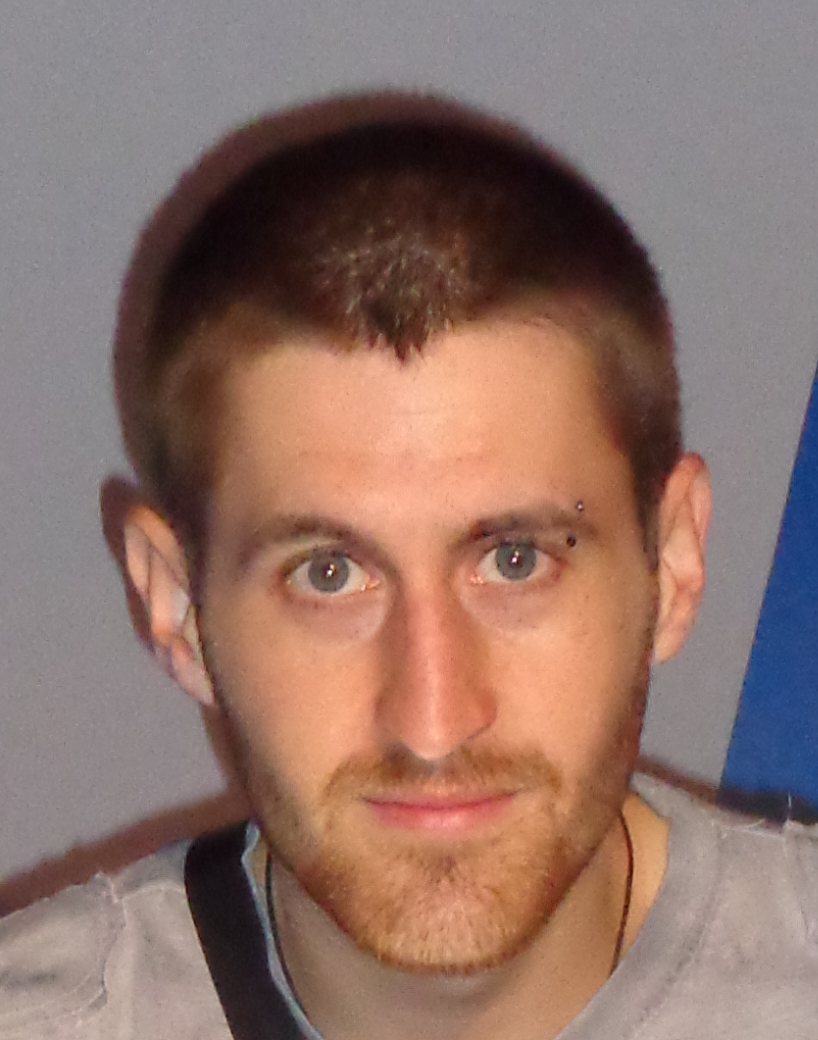}}]
{Patrice Camelin}  made a BTS Génie Optique Photonique before continuing
his studies at Université de Bourgogne (UB) were he obtained his License
degree in Physics and His Master degree in Physique Lasers et Matériaux.
He is currently a PhD student at the University of Nice Sophia-Antipolis.
His research interests include laser dynamics, passive mode-locking
and spatially extended systems.  
\end{IEEEbiography}

\begin{IEEEbiography}[{\includegraphics[width=2.5cm]{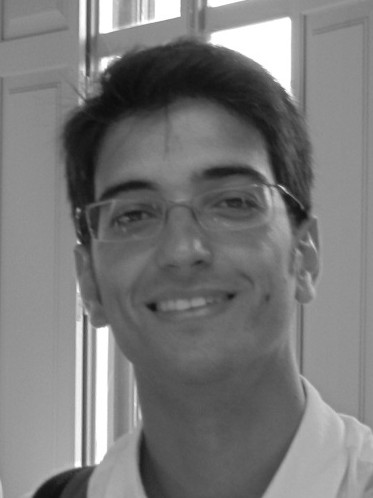}}]
{Daniel Chaparro González}  holds the Licenciado degree by the University
of Seville. He studied the postgraduate in Germany in the group of
Non-linear Photonics of the Institute of Applied Physics of the University
of Münster, where he worked on laser applications for optical tweezers
and manipulation and structuration of fluids. He wrote a thesis entitled
\textquotedblleft{}Fabrication of polymer diffraction gratings by
means of optically induced dielectrophoresis\textquotedblright{} with
which he obtained the Master of Science degree by the University of
Münster. Nowadays, he is a PhD student of the group of Non-linear
Waves of the University of the Balearic Islands working in theoretical
and experimental aspects of the use of semiconductor lasers to perform
remote measurements of distances. 
\end{IEEEbiography}

\begin{IEEEbiography}[{\includegraphics[width=2.5cm]{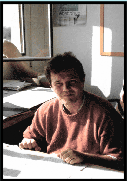}}]
{Salvador Balle}  (M\textquoteright{}92) was born in Manacor, Mallorca.
He graduated in Physics at the Universitat Autònoma de Barcelona,
where he obtained a PhD in Physics on the electronic structure of
strongly correlated Fermi liquids. After postdoctoral stages in Palma
de Mallorca and Philadelphia where he became interested in stochastic
processes and Laser dynamics, he joined in 1994 the Physics Department
of the Universitat de les Illes Balears, where he is Professor of
Optics since 2006. His research interests include laser dynamics,
semiconductor optical response modeling, multiple phase fluid dynamics
and laser ablation. 
\end{IEEEbiography}

\begin{IEEEbiography}[{\includegraphics[width=2.5cm]{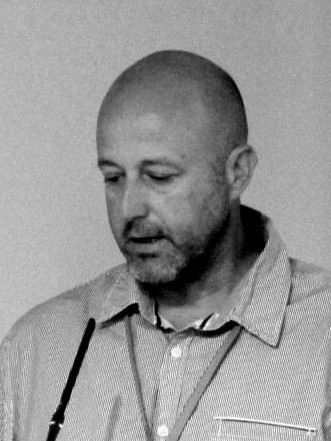}}]
{Massimo Giudici }  (M\textquoteright{}09) received the \textquotedbl{}Laurea
in Fisica\textquotedbl{} from University of Milan in 1995 and Ph.D
from Université de Nice Sophia-Antipolis in 1999. He is at present
full professor at Université de Nice Sophia-Antipolis and deputy director
of the laboratory \textquotedblleft{}Institut Non Linéaire de Nice\textquotedblright{},
where he carries out his research activity. Prof. Giudici's research
interests revolve around the spatio-temporal dynamics of semiconductor
lasers. In particular, he is actively working in the field of dissipative
solitons in these lasers. His most important contributions concerned
cavity solitons in VCSELs, longitudinal modes dynamics, excitability
and stochastic resonances in semiconductor lasers and the analysis
of lasers with optical feedback. \end{IEEEbiography}

\end{document}